\documentclass[useAMS,usenatbib]{mn2e}
\usepackage[dvips]{graphicx}

\title[The Dusty Aftermath of the V1309 Sco Binary Merger]{The Dusty Aftermath of the V1309 Sco Binary Merger}
\author[C. P. Nicholls et al.]{C. P. Nicholls$^1$\thanks{E-mail: cnicholls@physics.ucsd.edu}, C. Melis$^1$, I. Soszy{\'n}ski$^2$, A. Udalski$^2$, M.\,K. Szyma{\'n}ski$^2$, \and M. Kubiak$^2$, G. Pietrzy{\'n}ski$^{2,3}$, R. Poleski$^{2,4}$, K. Ulaczyk$^2$, {\L}. Wyrzykowski$^2$, \and S. Koz{\l}owski$^2$, and P. Pietrukowicz$^2$\\
$^1$Center for Astrophysics and Space Science, University of California San Diego, La Jolla, CA 92093, USA\\
$^2$Warsaw University Observatory, Al.~Ujazdowskie~4, 00-478~Warszawa, Poland\\
$^3$Universidad de Concepci{\'o}n, Departamento de Astronom{\'i}a, Casilla 160--C, Concepci{\'o}n, Chile\\
$^4$Department of Astronomy, Ohio State University, 140 W. 18th Ave., Columbus, OH 43210, USA}

\begin{document}
\setlength{\parskip}{0mm}
\date{Accepted 2013 January 6.  Received 2013 January 4; in original form 2012 October 31.}

\pagerange{\pageref{firstpage}--\pageref{lastpage}} \pubyear{2013}

\maketitle

\label{firstpage}

\begin{abstract}
\vspace{-2mm}

We present mid-IR photometry and spectra of the merged binary V1309 Sco taken between 18 and 23 months after  outburst. Strong mid-IR emission and a solid state absorption feature indicate the presence of a significant amount of dust in the circumstellar environment. The absence of detectable mid-IR emission before the outburst suggests this dust was produced in the eruptive merger event.

Model fits to the solid state absorption feature constrain the constituent species and column density of the dust around V1309 Sco. We find the absorption feature can be reproduced by large (3\,$\micron$) amorphous pyroxene grains at a temperature of 800\,K. This grain size, if confirmed with longer wavelength spectroscopy and modelling, would be suggestive of dust processing in the circumstellar environment. The data in hand do not allow us to discriminate between disk or shell configurations for the dusty material.

\end{abstract}

\begin{keywords}
stars: individual: V1309 Sco -- stars: circumstellar matter -- binaries: close 
\end{keywords}

\section{Introduction}

V1309 Sco is the first documented case of a binary merger \citep{tylenda}. Discovered as Nova Sco 2008 when it erupted in September 2008 \citep{nakano}, its subsequent evolution marked it as a new member of the `red nova' class of stellar outbursts, with a spectrum rapidly evolving from F to M spectral types (Rudy et al.\ 2008a, 2008b) \nocite{rudy08a,rudy08b} and line profiles transitioning from absorption to emission \citep{mason}. The red novae were proposed by \cite{sokertylenda} to be binary mergers, and include the equally enigmatic objects V838 Mon \citep{tylendaetal05} and V4332 Sgr \citep{kaminski}.

\cite{tylenda} presented the crucial evidence that revealed the nature of V1309 Sco, and possibly the class to which it belongs. They showed that the photometric variability found in its OGLE $I$-band light curve clearly marks V1309 Sco's progenitor as an early K giant contact binary with an orbital period of $\sim$\,1.4 days. This orbital period decreased in the 6 years prior to eruption, simultaneous with a slow increase of the system brightness, and they found no evidence of periodicity in V1309 Sco's light curve after its eruption. They calculated that the energy released in the outburst was consistent with the dissipation of orbital energy expected for a merger between the progenitor's components. 

Assuming V1309 Sco's outburst was indeed the result of a stellar merger, questions remain as to the nature of the remnant. In many binary models a phase of deep contact is reached as the orbit shrinks, resulting in mass and angular momentum loss from the system. This phase may lead to the formation of an excretion disk, which could remain around the coalesced star after the outburst \citep{webbink76,ibenlivio,shu79}. Mid-IR measurements of the remnant are an ideal way to probe dust production during the merger process.

In this Letter we show that V1309 Sco has become dominated by mid-IR emission since eruption. This can be attributed to dusty circumstellar material that likely formed during the merger, and we comment on the dust species that may constitute this matter. We also present evidence of dust processing.

\vspace{-9mm}
\section{Observations}

\begin{figure*}
\begin{center}
\includegraphics[width=0.9\textwidth]{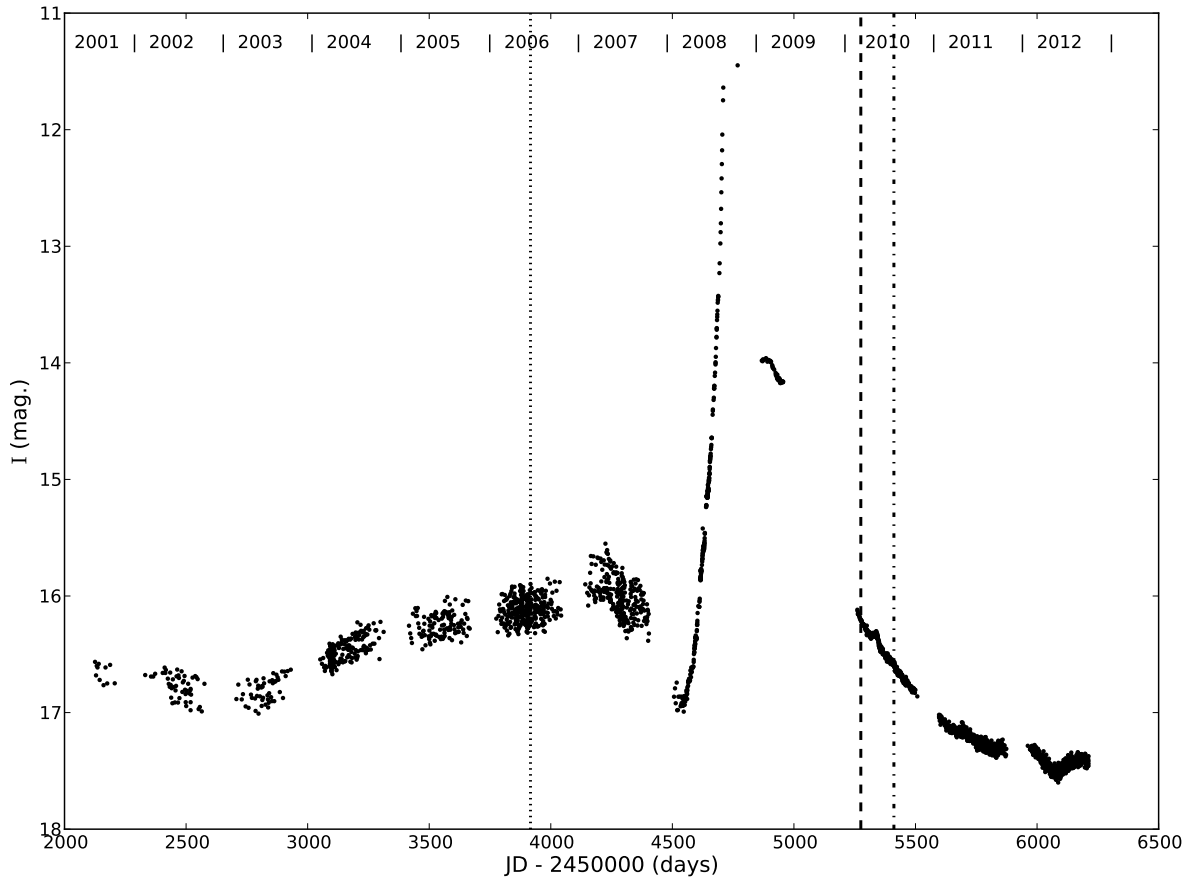}
\end{center}
\caption[]{OGLE $I$-band light curve for V1309 Sco. The peak luminosity is not covered as the object was overexposed in the OGLE images. Gaps denote times when the object was too close to the Sun to observe. The dashed line shows the epoch of the WISE observations, and the dot-dashed line shows the epoch of the TReCS observations. The dotted line shows the approximate epoch of the AKARI non-detection; see Section~\ref{dust}.}
\label{lightcurve}
\end{figure*}

The Optical Gravitational Lensing Experiment \citep[OGLE;][]{ogle} provides long-term optical light curves of many sources in the Galactic centre and the Magellanic clouds. V1309 Sco was monitored during the OGLE-III and OGLE-IV phases of the project, resulting in thousands of regular photometric measurements spanning more than ten years.
The OGLE III + IV $I$-band light curve of V1309 Sco is shown in Fig.~\ref{lightcurve}. Here we include new photometry from the 2011 and 2012 observing seasons of the OGLE-IV project that was not presented in \cite{tylenda}.

The Wide-field Infrared Survey Explorer \citep[WISE;][]{wisemn} satellite surveyed the entire sky in four mid-IR photometric bands centred on 3.4, 4.6, 12, and 22\,$\micron$. V1309 Sco was observed on 2010 March 19, with detections in all four bands. We downloaded the magnitudes from the NASA/IPAC Infrared Science Archive and converted them to fluxes in Jansky using the formula in \cite{wisemn}, and applied colour corrections appropriate for a 400$\,$K blackbody (see Section~\ref{dust}). We do not report the more recent WISE 3-band Cryo Release photometry of V1309 Sco as its reliability is unclear at this time.

V1309 Sco was observed by the Thermal-Region Camera Spectrograph (TReCS) at Gemini South under program GS-2010B-C-7 on 2010 August 3, as part of a project to characterise dust forming novae. We downloaded the raw N-band spectrum and narrow band images from the Gemini Science Archive. We extracted photometry of V1309 Sco using standard \textit{IRAF} procedures and converted the magnitudes to fluxes in Jansky. We assume 10 per cent error bars on these measurements. We reduced the mid-IR spectrum using the Gemini \textit{IRAF} mid-IR spectral reduction pipeline, \textit{msreduce}, and flux calibrated by shifting it vertically to match the flux levels of the narrow-band photometry.

\begin{figure*}
\begin{center}
\includegraphics[width=0.9\textwidth]{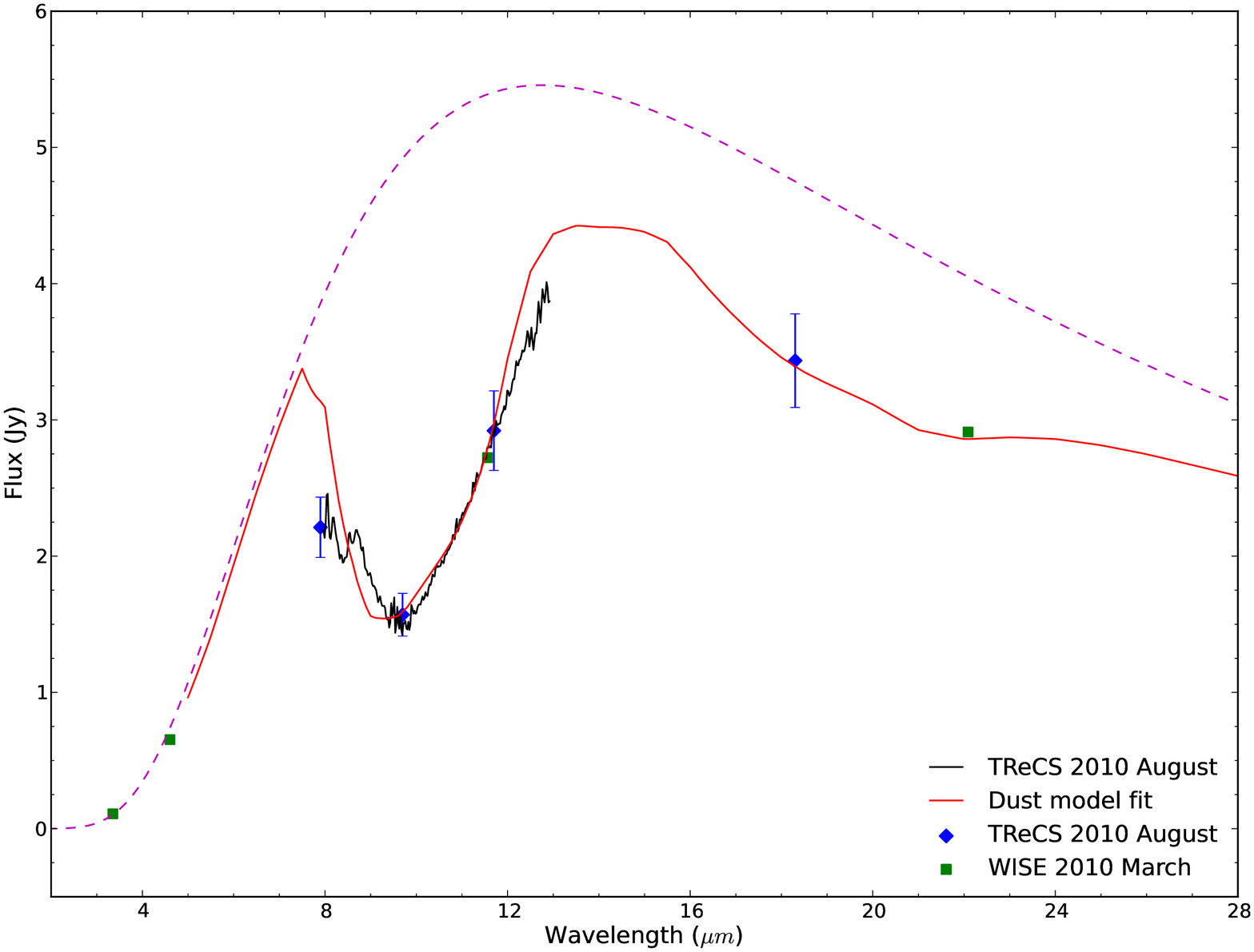}
\end{center}
\caption[]{Mid-IR SED of V1309 Sco. The TReCS spectrum is shown by the black line, the WISE photometry by the green squares and the TReCS photometry by the blue diamonds. The red line represents our simple dust model fit. The dashed purple line is a blackbody fit to the continuum at a temperature of 400\,K. The WISE error bars are comparable to the point size.}
\label{mir}
\end{figure*}

The WISE and TReCS mid-IR photometry is shown in Table~\ref{obstable}. The photometry and the TReCS spectrum are plotted in Fig~\ref{mir}.

\begin{table}
\centering
\caption{Mid infrared photometry of V1309 Sco}
\label{obstable}
\begin{tabular}{lrrrrr}
\hline
\multicolumn{1}{c}{JD}  &  \multicolumn{1}{c}{Source}  &  \multicolumn{1}{c}{Filter}  &  \multicolumn{1}{c}{$\lambda$}  &  \multicolumn{1}{c}{Flux}  &  \multicolumn{1}{c}{Error}\\
 & & & \multicolumn{1}{c}{($\micron$)} & \multicolumn{1}{c}{(Jy)} & \multicolumn{1}{c}{(Jy)}\\
\hline                                                                           
2455275.14 & WISE & W1 & 3.35 & 0.109 & 0.002  \\
2455275.14 & WISE & W2 & 4.60 & 0.653 & 0.013  \\
2455275.14 & WISE & W3 & 11.6 & 2.723 & 0.028  \\
2455275.14 & WISE & W4 & 22.1 & 2.911 & 0.035  \\
2455411.22 & TReCS & Si1 & 7.73 & 2.211 & 0.221  \\
2455411.22 & TReCS & Si3 & 9.69 & 1.570 & 0.157  \\
2455411.22 & TReCS & Si5 & 11.7 & 2.921 & 0.292  \\
2455411.22 & TReCS & Qa & 18.3 & 3.435 & 0.344  \\
\hline
 \end{tabular}
 \end{table}

\vspace{-9mm}
\section{Results and Discussion}

\subsection{Optical Light Curve}

The features that \cite{tylenda} noted in the OGLE $I$-band light curve, namely the slow brightening of the progenitor, the fading prior to eruption, the smooth outburst beginning $\sim 6$ months before its discovery as Nova Sco 2008, and the decline to magnitudes fainter and with smaller scatter than the progenitor, can be clearly seen in Fig.~\ref{lightcurve}. With the addition of more recent photometry, it is clear that V1309 Sco for the most part continues to decline in brightness, opposite behaviour to that shown by red novae V838 Mon and V4332 Sgr \citep{tylenda05,tylendaetal05}. It remains to be seen whether the brightening starting at JD $\sim$ 2456100 days is periodic or the beginning of a reversal. There also appears to be a slight increase in scatter over the two most recent observing seasons.

Using \textit{IRAF}'s Phase Dispersion Minimisation tool \citep{stellingwerf}, we find no evidence of periodic variability in the new photometry. This lack of periodicity indicates that the variability of the remnant is unlikely to be pulsation or binary-induced variations. It is unclear at this stage whether the post-outburst scatter and faintness is an intrinsic result of the merger. We discuss an alternative scenario in Section~\ref{nature}.

\vspace{-5mm}
\subsection{Mid-Infrared Measurements}
\label{dust}

We searched for previous mid-IR measurements of V1309 Sco using its WISE coordinates of 17:57:32.94, -30:43:09.84 which are accurate to $0\farcs043$ in RA and $0\farcs044$ in Dec. No detection was found at this position in the IRAS or AKARI catalogues. At 18\,$\micron$, AKARI is 100 per cent complete down to 0.3\,Jy (Ishihara et al.\ 2010), \nocite{akarimn} so at that epoch (2006) V1309 Sco was fainter than 0.3\,Jy at 18\,$\micron$, a difference of more than 9 times the 2010 TReCS Qa-band measurement (Table~\ref{obstable}). We note that AKARI detected a bright source (2.7\,Jy) in this band 2 arcmin away from the position of V1309 Sco, further suggesting AKARI should have been capable of detecting V1309 Sco if it was as bright as it was in 2010. At 12\,$\micron$, IRAS is estimated to be 95 per cent complete down to 0.46\,Jy\footnote{IRAS Explanatory Supplement, http://irsa.ipac.caltech.edu/IRASdocs/exp.sup/}. From its non-detection at the epoch of IRAS (1983), V1309 Sco was likely fainter than 0.46\,Jy at 12\,$\micron$, a difference of more than 6 times the 2010 TReCS Si5 flux.

Although V1309 Sco was undetected in the mid-IR before its eruption, as of 2010 it had reached flux densities of over 3\,Jy at some mid-IR wavelengths (as seen in Table~\ref{obstable} and Fig.~\ref{mir}), despite the optical flux having declined to pre-outburst values. This suggests that the merger between the progenitor's binary components produced not only a huge optical flare but also a significant amount of dust. The pre-burst fading of the system (visible at JD $\sim$ 2454500 days in Fig.~\ref{lightcurve}) was proposed by \cite{tylenda} to signal the formation of a dusty excretion disk. The recent high mid-IR flux certainly supports the hypothesis that the binary merger triggered significant dust formation, and that this dust has been retained in the circumstellar environment of the system.

To learn more about the dust surrounding V1309 Sco we made a simple model fit to the mid-IR measurements. The model takes the form of
\begin{equation}
D = f \left[ B_{\nu, cont} - \sum_{i=1}^{n} \left(a_{i} A_{i} B_{\nu, i} \right) \right]
\label{model}
\end{equation}
where $D$ is the dust model, $f$ is a constant scaling the total received flux, $B_{\nu, cont}$ is a blackbody representing the dust continuum emission, $a_{i}$ is a constant scaling the amount of the $i$th absorbing species, $A_{i}$ are the absorption coefficients for the $i$th absorbing species, and $B_{\nu, i}$ is a blackbody at the temperature of $i$th absorbing species. Due to the sparseness of the data we allow only a single dust continuum, and one absorbing component.

We used absorption coefficients calculated by \cite{min}, using both Gaussian Random Field (GRF) and Distribution of Hollow Spheres (DHS) grain shape distributions. Our dust species grid included amorphous olivine and pyroxene, amorphous Mg-rich olivine and pyroxene, crystalline olivine and pyroxene, and silica. For each species various grain sizes between 0.1 and 4.0$\, \micron$ were considered. Our temperature grid covered $T=200\,$K to $1500\,$K in 50\,K steps. We determined the best fit to the data by minimising $\chi^2$.

We find the dust continuum is best fit with a temperature of 400\,K. Of the features in V1309 Sco's mid-IR spectrum, most notable is the broad absorption band at $9.7\, \micron$, usually attributed to amorphous silicates. We find the best fit to the absorption features is amorphous pyroxene $\rm{MgFeSi_2O_6}$ grains 3.0$\,\micron$ in size, at 800\,K, with a DHS shape distribution. In general we found that $\chi^2$ was more sensitive to grain size than to species, temperature, or shape, with larger grain sizes (3.0 and 4.0$\, \micron$) corresponding to lower $\chi^2$ values, reproducing the broad 9.7$\,\micron$ feature with greater accuracy. The best fit is shown in Fig~\ref{mir}.

It is clear that even our best fit is somewhat lacking. This poorness of fit is likely due to the simplistic nature of our model, however a more complete SED is required before detailed modelling is undertaken. Although we quote the best fit species, temperature, and shape distribution as likely properties of V1309 Sco's circumstellar dust, we note here that it is difficult to properly constrain these with the data in hand. However we are confident that the broad $9.7\, \micron$ feature indicates large grains.

Because V1309 Sco's mid-IR SED shows dust absorption instead of emission features, indicating an optically thick circumstellar environment, we are able to derive only the column density of the dusty circumstellar matter, not the total dust mass. Column density may be calculated by dividing the optical depth by the extinction cross section of the absorbing grains. The optical depth at a particular wavelength can be calculated from the observed flux and the continuum model fit; the extinction cross section is $\sigma_{\lambda} = Q_{\lambda} \pi a^2$, where $Q_{\lambda}$ is the extinction coefficient of that particular grain type and size at a particular wavelength, and $a$ is the grain radius. We used extinction coefficients for `astronomical silicate' at $\lambda = 9.7 \micron$ and $a = 1.5 \micron$ calculated by \cite{laordraine}, as extinction coefficients for our modelled dust grains were unavailable. We find that in the 9.7$\, \micron$ feature, the column density of the 3$\, \micron$ amorphous pyroxene grains is $4.3 \times 10^6\, \rm{cm^{-2}}$. 

As \cite{bairdcardelli} note, column density and total dust mass may not necessarily correlate, as an extended dust cloud may have a large mid-IR excess (indicating high mass) but low column density due to its large volume. Therefore without an estimate of the radial extent of the dust, we are unable to estimate its mass. We stress that the column density derived here is a lower limit for the total dust column density of V1309 Sco's circumstellar environment, as it includes only the 3$\, \micron$ absorbing pyroxene dust grains, and neglects any particles that may contribute to the dust continuum emission. 
The details of V1309 Sco's circumstellar environment are shown in Table~\ref{cse}.

\begin{table}
\centering
\begin{minipage}{70mm}
\caption{The Circumstellar Environment of V1309 Sco}
\label{cse}
\begin{tabular}{lc}
\hline
Parameter  &  Value  \\
\hline                                                                           
$T_{\rm{cont}}$  &  400\,$\pm$25\,K \\
$T_{\rm{feat}}$  &  800\,$\pm$25\,K \\
Dust species  &  Amorphous pyroxene $\rm{MgFeSi_2O_6}$  \\
Grain size  &  $\sim 3.0\, \micron$  \\
Column density$^a$\footnotetext{$^a$ Calculated at the 9.7$\, \micron$ feature}  &  $(4.3 \pm 0.43) \times 10^6\, \rm{cm^{-2}}$  \\
Optical depth$^a$  &  0.71\,$\pm$\,0.07 \\
\hline
 \end{tabular}
 \end{minipage}
 \end{table}

Two particular results should be noted from our dust fitting. The first is that the warm temperature of the dust suggests  it formed recently, consistent with the absence of a previous mid-IR detection. The second is that the large size of the grains producing the absorption features suggests the dust has been processed in the circumstellar environment: dust formed in the atmospheres of AGB stars and returned to the ISM is typically sub-micron in size, and larger grains require processing in a stable environment. It is difficult to say at present whether V1309 Sco's dusty shroud is a disk or a shell, but the processing of dust grains to sizes of a few microns certainly implies a quick evolution, if we assume the dust was formed as part of the recent merger event.

\vspace{-8mm}
\subsection{The Nature of V1309 Sco}
\label{nature}

From the work of \cite{tylenda} and the results presented here, it seems that V1309 Sco was a contact binary that produced a nova-like outburst as the two stars merged, and is now surrounded by warm optically thick dusty matter showing strong absorption from processed amorphous pyroxene particles. The lack of periodicity in the post-outburst lightcurve supports the hypothesis that V1309 Sco is now a single star. It seems likely that the dusty envelope formed during the merger.

Many models of binary interaction involve mass loss and formation of a disk, such as the circumstellar excretion disk proposed by \cite{webbink76}. Because V1309 Sco's progenitor was a photometrically varying contact binary, we are viewing the system near its equatorial plane, but not necessarily exactly edge on \citep{eccentricpaper}. It is possible that the large amount of dust we detect resides in a disk seen slightly tilted from edge-on, meaning although it produces the large mid-IR excess we observe it does not completely enshroud the system, extinguishing only part of the optical flux. This could account for the recent faint $I$-band magnitudes of V1309 Sco as seen in Fig.~\ref{lightcurve}, and the pre-outburst fading. The post-outburst irregular small variations seen in the light curve may also be caused by a slightly tilted disk, as clumps of dust pass in front of the line of sight. However, with our current data we are unable to discern whether V1309 Sco's dust is in a disk or another configuration.

A similar mid-IR excess was recently observed in V4332 Sgr, a red nova that erupted in 1994 \citep{banerjee,kaminski}. It too shows a large 9.7\,$\micron$ absorption feature indicating an optically thick oxygen rich dusty environment, and evidence of alumina. Both \cite{banerjee} (and references therein) and \cite{kaminski} proposed that V4332 Sgr's dust occupies a stable disk that is probably edge on to our line of sight and hence causing the observed high IR-to-optical luminosity ratio. No previous measurement of the column density of V4332 Sgr's dust has been made, but we can make a rough estimate if we assume the grains are similar to what we find for V1309 Sco. In this way, we estimate that the column density of the absorption feature seen toward V4332 Sgr is $\sim 5 \times 10^6\, \rm{cm^{-2}}$, comparable to the value we calculate for V1309 Sco. More detailed dust modelling would constrain this value further. 

V1309 Sco, as the first confirmed binary merger, provides an ideal laboratory to test merger models. Proposed merger remnants in the literature include blue stragglers, rapidly rotating giants like FK Com \citep{gazeasstepien}, and giants accreting from a circumstellar disk \citep{melis09}. It remains to be seen whether V1309 Sco shows properties indicative of these types of objects.

V1309 Sco is a most fascinating object and worthy of further study. Continued photometric monitoring is important to track its luminosity and variability evolution. Optical spectra are necessary to provide measurements of its rotational velocities. Mid-IR spectra at longer wavelengths are essential to determine the exact composition of its dusty circumstellar environment, and mid-IR monitoring will reveal any time evolution of this dust. Whether the dust surrounding V1309 Sco resides in a disk is, we feel, one of the most pressing questions.

\vspace{-7mm}
\section*{Acknowledgements}
We thank Michiel Min and Clio Gielen for providing us with dust absorption coefficients.
C.M. acknowledges support from the National Science Foundation under award No.\ AST-1003318.
The OGLE project has received funding from the European Research Council under the European Community's Seventh Framework Programme (FP7/2007-2013) / ERC grant agreement no. 246678 to AU.
Based on observations obtained at the Gemini Observatory (acquired through the Gemini Science Archive), which is operated by the Association of Universities for Research in Astronomy, Inc., under a cooperative agreement with the NSF on behalf of the Gemini partnership: the National Science Foundation (United States), the Science and Technology Facilities Council (United Kingdom), the National Research Council (Canada), CONICYT (Chile), the Australian Research Council (Australia), Ministério da Ciência, Tecnologia e Inovação (Brazil) and Ministerio de Ciencia, Tecnología e Innovación Productiva (Argentina). 
This publication makes use of data products from the Wide-field Infrared Survey Explorer, which is a joint project of the University of California, Los Angeles, and the Jet Propulsion Laboratory/California Institute of Technology, funded by the National Aeronautics and Space Administration.
This research has made use of the NASA/IPAC Infrared Science Archive, which is operated by the Jet Propulsion Laboratory, California Institute of Technology, under contract with the National Aeronautics and Space Administration.

\vspace{-6mm}
\bibliographystyle{mn2e}
\bibliography{bibliographynew}

\label{lastpage}

\end{document}